\documentclass[12pt]{article}
\usepackage{epsfig}
\usepackage[active]{srcltx}
\usepackage{graphicx}
\newcommand{\Si}{\mbox{\boldmath $\Sigma$}}
\newcommand{\bal}{\mbox{\small\boldmath $\alpha$}}
\newcommand{\hp}{\mbox{\boldmath$\hat{p}$}}

\newcommand{\E}{\mbox{\boldmath $E$}}

\newcommand{\eref}[1]{(\ref{#1})}

\newcommand{\A}{\mbox{\boldmath$A$}}

\newcommand{\Hb}{\mbox{\boldmath$H$}}

\newcommand\fr{\displaystyle\frac}
\newcommand\lt{\left}
\newcommand\rt{\right}
\newcommand{\sh}{Schr\"odinger equation}
\begin{document}
\large
\begin{center}
{\bf
A model of the ball lightning}\\ \vspace{0.2cm}
\normalsize

V.K.Ignatovich\\
{\it Laboratory of Neutron Physics, Joint Institute for Nuclear Research,
141980 Dubna Moscow region, Russia}\\
\vspace{0.4cm}

\parbox{125mm}{The ball lightning is supposed to be a shock wave
of a point explosion frozen with electrostriction forces of the
internal strong laser discharge. The life time of the ball with
modest parameters is calculated.} \vskip 0.5cm

\noindent
PACS numbers:52.80.Mg, 41.90.Jb, 92.60.Pw
\end{center}
\vspace{0.5cm}

\large

After many years I decided to rewrite the article about
electromagnetic model of the ball lightning, which was first
published in Russian in 1980~\cite{ig2}, and later in
1992~\cite{ig}.

Electromagnetic models of ball lightning were considered, for
instance, in papers~\cite{kap}-\cite{kul} though in~\cite{smi}, it
is said that they can not explain the observed parameters. Most of
the authors treated the ball lightning as a plasma formation. In
particular, in~\cite{zen} the ball lightning was supposed to be a
cavity with radiation locked by surrounding plasma.

One similar model was also considered in~\cite{ig2}. In that model
the photons were contained in a sphere surrounded by a gas with
anomalous dispersion at frequency of the stored radiation. There
and in~\cite{ig} it was also considered another model which can be
visualized as a shock wave of a point explosion in the atmosphere.
This shock wave can be stopped and frozen by a powerful laser
discharge behind the shock wave front, if it arises in whispering
gallery mode. The stopped shock  represents a thin spherical shell
filled with electromagnetic radiation. Radiation is retained in
the shell because of the total internal reflection, and the shell
is kept integer because of electrostriction forces and surface
tension created by the radiation.

This model can explain both the high energy and the long life time
of the ball lightning. More over it throws some light on the
nature of other phenomena, such as hurricane and tornado and can
be used in many other branches of physics. In~\cite{tor} it is
also described how such a model can explain unusual properties of
the ball lightning, its ability to penetrate indoors through slits
and windowpanes, and to move against wind.

To show how the model works we use an analogy with quantum
mechanics of a particle. Since the stationary equation for
electric field $E$ (for simplicity we use here the scalar
approximation)
\begin{equation}
[\Delta +n^2k^2_0]E({\bf r})=0, \label{ab1}
\end{equation}
where $k_0=\omega/c$, $\omega$ is the photon frequency, $c$ is the
light speed in vacuum and $n$ is the index of refraction, is very
similar to the \sh
\begin{equation}
[\Delta +k^2-u({\bf r})]\psi({\bf r})=0,
\label{aa1}
\end{equation}
for particles~\cite{land}. Here $\psi$ is the wave function of a particle
with wave number $k^2$, $u$ is a potential energy measured in units
$\hbar^2/2m$ and $m$ is the mass of the particle.

To transform equation
(\ref{ab1}) to the form (\ref{aa1}) it is necessary only to denote
\begin{equation}\label{n1}
u=(1-n^2)k^2_0.
\end{equation}
For the equation (\ref{aa1}) it is known that, if $u$ is a
potential well, the particle can have a bound state inside it. The
same can be said about the photon. If the potential well is inside
a matter, and the well depth increases with increase of the matter
density, then the particle in a bound state in such a well is not
only kept by the well but also compresses the matter. This is the
analog of electrostriction forces.

Let us illustrate such an effect with the help of a slow neutron.
It is known (see, for example~\cite{ig1}) that interaction of
neutrons with nuclei creates neutron-matter interaction potential
$u=4\pi N_0b$, where $N_0$ is the number of atoms in a unit
volume, and $b$ is the coherent scattering amplitude. If $b>0$
then $u>0$ and matter repels the neutron. However if $b<0$, the
matter attracts the neutron. So a substance with a negative $u$ is
a potential well for neutrons. But, and it is very important,
because $u$ is proportional to atomic density, not only matter
holds neutron in the well, but the neutron itself, if it is in a
ground state, also holds the matter, i.e., it resists expansion of
the matter.

Indeed, it is shown on fig.1 that, when the matter expands, the
density $N_0$ and the well depth become lower. It leads to rise of
the lowest energy level. Therefore expansion requires some work to
increase the neutron energy.
\begin{figure}[h!]
\setlength{\unitlength}{1mm}\thicklines
\begin{picture}(120,60)
\put(0,0){\makebox(120,10){Fig.1}}
\put(15,15){\line(1,0){30}}
\put(45,15){\line(0,1){40}}
\put(15,15){\line(0,1){40}}
\multiput(15,20)(6,0){5}{\line(1,0){3}}
\put(5,55){\makebox(5,5){$U$}}
\put(5,20){\makebox(5,5){$E$}}
\put(75,25){\line(1,0){48}}
\put(123,25){\line(0,1){30}}
\put(75,25){\line(0,1){30}}
\multiput(75,28)(6,0){8}{\line(1,0){3}}
\put(62,55){\makebox(5,5){$U-dU$}}
\put(62,28){\makebox(5,5){$E+dE$}}
\put(15,10){\makebox(30,5){$r$}}
\put(75,20){\makebox(48,5){$r+dr$}}
\end{picture}
\caption{\label{f1} Illustration of striction of matter by a
particle in the ground state in a well created by the matter.}
\end{figure}

It is easy to calculate the force with which the particle in the
ground state compresses the matter. The equation for bound state
energy can be represented as
\begin{equation}\label{n2}
\exp(2ikr)\rho^2(k)=1,
\end{equation}
where $k=\sqrt{2mE/\hbar^2}$, $m$ is the particle mass, and the
bound state energy $E>0$ is counted from the bottom of the well.
This \eref{n2} is a condition for stationarity. It means, that if
wave function of the particle in the well moving to the right has
some value near the right wall, then after reflection with
amplitude $\rho(k)$ from that wall, propagation to the left,
reflection with the same amplitude (we assume the well to be
symmetric) from the left wall, and coming back to the right wall
should have the same initial value of its wave function.
Reflection amplitude is
\begin{equation}\label{n3}
\rho(k)=\fr{k-ik'}{k+ik'}=\exp(-2i\varphi),\qquad
\varphi=\arccos(k/\sqrt u),
\end{equation}
where $k'=\sqrt{u-k^2}$ and $u\propto1/r$ determines the depth of
the well. Equation \eref{n2} for the ground state is reduced to
\begin{equation}\label{n5}
kr-\arccos(k/\sqrt u)=0,\qquad \cos(kr)=k/\sqrt u,
\end{equation}
and it is a simple exercise to calculate the force $dE/dr$ from
this equation. The force is negative, which means compression of
the substance.

Of course, the striction force created by a single particle is
small. But if the number of particles large, the force can be also
large. This may happen to be very important for,say, neutron
stars. In a neutron star every neutron has a potential $u=4\pi
N_0b$, where $N_0$ here is the density of neutrons. The
compression energy here is proportional to $N_0^2$. It is very
high and may be even greater than the gravitational one. This
question is considered in~\cite{nst}

In the paper~\cite{kul} the analogous interaction was considered
for electrons. It is shown, that at some density of plasma
exchange attractive interaction can become larger than direct
Coulomb repulsion, and it leads to the coherent binding of plasma
particles.

All these considerations can be applied also to $\gamma$-quanta.
And it leads to our model for the ball lightning.
Indeed $n^2=\epsilon=1+4\pi N_0\alpha,$
where $N_0$ is the number of molecules in a unit volume and
$\alpha$ is polarizability of a molecule. So, the potential
\begin{equation}\label{m1}
u=(1-n^2)k^2_0=-4\pi N_0\alpha k^2_0,
\end{equation}
is negative for positive $\alpha$. It means that usually matter
attracts photons, and this attraction is known as "pondermotive
force" and self focussing. But in analogy with quantum mechanics
of a particle we can speak also about bound levels of photon, and
such approach will give us a possibility to estimate the life time
of the ball.

In the case of anomalous dispersion or at high frequencies the
permittivity $\epsilon$ can be less then unity. In these cases
$\alpha<0$, and matter repels photons. This happens, for instance
in the case of interaction with plasma.

Interaction of atoms and molecules with electromagnetic field is
described by the expression
\begin{equation}\label{m2}
U_1=-dE=-\alpha E^2=-4\pi\alpha N_{\gamma}\hbar\omega,
\end{equation}
where $d=\alpha E$ is the induced dipole moment (we suppose the
absence of own dipole moment), $\alpha$ is the polarizability of
the molecule and $N_{\gamma}$ is the number density of photons.
This interaction shows, that the forces drawing matter inside the
photon field is proportional to gradient of its density.

From quantum mechanics it follows~\cite{lan,jac}, that
\begin{equation}\label{m3}
\alpha=\frac{e^2}{2m_e}\sum_{k\ne
0}\frac{f_{0k}}{\omega^2_{0k}-\omega^2-\imath\omega\Gamma_k},
\end{equation}
where $\omega$ is the incident photon frequency,  $e$ is the
electric charge, $m_e$ is the mass of the electron,
$\omega_{kl}=\omega_k-\omega_l$ are the eigen frequencies defined
by transition $k\rightarrow l$, $\omega_k$ are energy levels of
the electron in the atom, $\Gamma_k$ is the width of the
transition, $f_{lk}$  are oscillator strengths given by
\begin{equation}\label{m4}
f_{kl}=(2m_e/\hbar^2)\hbar\omega_{lk}|d_{kl}|^2,
\end{equation}
and $d_{kl}=<k|r|l>$ is a matrix element of dipole transition
between states $|l>$ and $|k>$.

For $\omega<\omega_{01}$ an unexcited atom is pulled into regions
of space with larger density of photons and, if it is excited, it
will with the highest probability emit the same photon, which is
already present in the field, because the matrix element of the
transition is proportional to the square root of the total number
of the photons present in the mode. So, the photons in the intense
coherent field can not be scattered and atoms are
stabilized~\cite{fed}-\cite{kulan}.

This happens to neutral molecules. But electrons and ions are repelled by
electromagnetic field, because their eigen frequencies in space can be put
equal to zero. It means, that if ionization happened at the explosion
then the light electrons will fly before the shock wave front, and after
creation of thin spherical layer filled with intense electromagnetic field
electrons remain to be separated from ions left behind the shock wave, and
we obtain a charged spherical capacity.

Let us consider parameters of the ball lightning. We shall take
radius and the energy of it to be given and equal to 10 cm and 10 kJ
respectively and then estimate its life time.

In the spherical coordinate system the equation for radial part of the
electromagnetic potential
\begin{equation}\label{1n}
A({\bf r})=\frac{R_L(r)}{r}P_L(\theta)\exp(\imath m\phi),
\end{equation}
looks like
\begin{equation}\label{2n}
\lt[\frac{d^2}{dr^2}+k^2-u(r)-\frac{L(L+1)}{r^2}\rt]R_L(r)=0,
\end{equation}
where $L$ is an orbital momentum of photons, or (\ref{2n}) can be
represented in the form
\begin{equation}\label{3n}
\lt[\frac{d^2}{dr^2}+k^2-V(r)\rt]R_L(r)=0,
\end{equation}
\begin{figure}[h!]
{\par\centering\resizebox*{12cm}{!}{\includegraphics{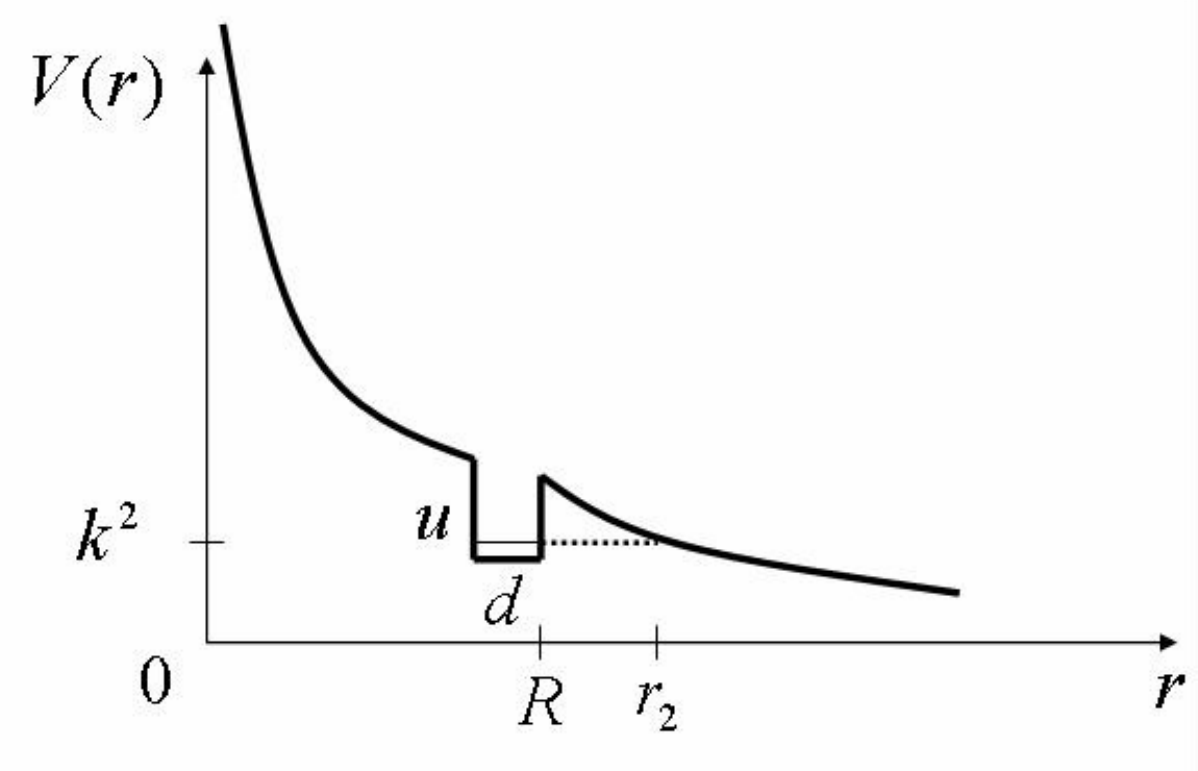}}\par}
\caption{\label{f3} The effective potential $V(r)$ in Eq.
(\ref{2n}).}
\end{figure}
with the effective potential $V(r)=L(L+1)/r^2+u(r)$ Fig. \ref{f3}.
The potential is positive, and $u$ represents a "pocket" on a
monotonously decreasing centrifugal potential curve. The photons
have a metastable state in this pocket. Since the scattering is
prohibited, the only way photons can leave this pocket is through
tunnelling.

Let the wave length of trapped radiation be $\lambda=10^{-4}$ cm.
Then, for sphere of radius $r_0=10$ cm we get $L=kr/2\pi\approx
10^5$. The life time $T$ can be estimated by expression $T=t_f/P$,
where $t_f$ is free flight time between two collisions with the
shock wave front and $P$ is the probability of tunnelling through
the potential barrier. Since $t_f<10^{-10}$ s, the probability $P$
must be very low. Let us find $P$ with the usual quasiclassical
approximation of quantum mechanics.
\begin{equation}
P=\exp(-2\gamma), \qquad
\gamma=\int_{R}^{r_2}\sqrt{L^2/r^2-k^2}\,dr. \label{a91a}
\end{equation}
The integration limits are determined by the relations
\begin{equation}\label{5n}
(L/R)^2=k^2+|u|,\qquad (L/r_2)^2=k^2.
\end{equation}
At large $L$ and small $u$ the integral in (\ref{a91a}) can be
approximated by the expression
\begin{equation}\label{6n}
\gamma=\int_0^{x_2}\sqrt{|u|-2L^2x/R^3}\,dx=
\frac13\left(\frac{|u|}{k^2}\right)^{3/2}L,
\end{equation}
where $x_2=r_2-R$. To get lifetime near $10^4$ s it is necessary
to have $\gamma\approx 20$ and for $L=10^5$ the value of
$|u|/k^2=\epsilon-1$ should be $\approx 10^{-2}$. It gives the
magnitude of the refraction index $\epsilon$ inside the shock wave
to be of the order of 1.007.

The angle $\phi$ of total reflection is defined from
$\sin\phi=1/n$. It shows that the width of the photon layer is
\begin{equation}\label{m6}
d=r_0[1-\sin\phi]\approx 0.01r_0\approx 0.1{\rm\ cm}.
\end{equation}
All these parameters are not extraordinary, so the life time of
order 10 000 seconds seems to be quite achievable.

If total energy is concentrated in photons, then the layer must
contain $10^{23}$ photons of energy 1 eV each. The density in the
layer then is equal $N_\gamma\approx 10^{27}$ m$^{-3}$. At such a
density the surface tension is $\sigma=(\epsilon-1)\hbar\omega
N_\gamma d\approx 10^3$ J/m$^2$. The surface tension is defined as
$\sigma=-dE_s/dS$, where $E_s$ is the total energy of the layer,
which here is $E_s=(1-\epsilon)\hbar\omega N_\gamma Sd$, and $S$
is the surface of the sphere.

Such a surface tension creates compressing pressure
$p=2\sigma/r=2\cdot10^4$ J/m$^3$. Since the normal atmospheric
pressure is of the order $10^5$ J/m$^3$, the gas density inside
the ball to withstand the compression should be only 20\% higher
than outside pressure, or the temperature inside gas should be
only 60 K higher than outside. Because of higher density inside
the photon film, the ball is heavier than environment and falls
down. If the gas density in the ball is lower than outside, its
temperature must be higher, and the ball can be lighter than the
air.

For photon frequencies very close to a resonance the magnitude of
$n^2-1$ can be higher, and therefore the higher will be the
surface tension and gas temperature inside the ball. Situation
improves even more, if one takes into account the Lorenz-Lorentz
correction.

The ball can be also a charged spherical capacitor, if during the
point explosion a separation of charges takes place. For instance
the light electrons will fly faster and become before the shock
wave front, while heavier positive ions remain behind it. The
energy of the capacitor depends on its charge. Let us suppose that
the charge is equal to $Q$. An outside electron is attracted by
the charge of ions with the force $F_q=Ee=9\times
10^9Qe/r_0^2=Q\times 10^{-7}$ N. But the photons repel it. The
interaction energy of an electron with the photon layer is
\begin{equation}
u_e=\pi(e^2/mc^2)\lambda^2\hbar\omega N_\gamma. \label{b91}
\end{equation}
Let us show how to derive it with the help od the Dirac equation
\begin{equation}\label{d1}
\lt[\gamma\lt( p-\fr ec A\rt)-mc\rt]\psi=0,
\end{equation}
where  $A$ is a four dimensional vector potential: $A=(-\A,\Phi)$,
$\Phi$ is its scalar and $\A$ is its three dimensional vector
part.

We can transform \eref{d1} into equation of the second order
multiplying it from the left by
\begin{equation}\label{d2}
\lt[\gamma\lt( p-\fr ec A\rt)+mc\rt],
\end{equation}
which gives
\begin{equation}\label{d3}
\lt[\lt( p-\fr ec
A\rt)^2-m^2c^2-\fr{e\hbar}c[\Si\Hb-i\bal\E]\rt]\psi=0.
\end{equation}
We neglect $\Phi$ and average of \eref{d3} over fast oscillations
of the field. Then \eref{d3} is facilitated to
\begin{equation}\label{d4}
\lt[p_0^2-m^2c^2-\hp^2-\lt(\fr e{2c}\rt)^2 \A^2\rt]\psi=0.
\end{equation}
We divide it by $2m$ ($m$ is electron mass) and reduce \eref{d4}
to nonrelativistic form
\begin{equation}\label{d5}
\lt[-\fr{\hbar^2}{2m}k^2+\fr1{2m}\hp^2+\fr{e^2}{4mc^2}\A^2\rt]\psi=0.
\end{equation}
Since $\E=d\A/cdt$, then we can substitute
$\A^2=(c/\nu)^2\E^2=\lambda^2\E^2$, and use the energy density in
the form $E^2/4\pi=\hbar\omega N_\gamma$. As a result we obtain
\begin{equation}\label{d15}
\lt[-\fr{\hbar^2}{2m}\Delta+\pi\fr{e^2}{mc^2}\lambda^2\hbar\omega
N_\gamma-\fr{\hbar^2}{2m}k^2\rt]\psi=0,
\end{equation}
where potential energy is given by \eref{b91}.

The repulsive force is proportional to the gradient of
$N_{\gamma}$. The distribution of gamma quanta is determined by
the Bessel function $J_L(kr)$, so $dJ_L(kr)/dr\approx
(L/r_0)J_L(kr)$. It means that the force can be estimated as
$F_e=Lu_e/r_0$, or $F_e\approx 10^{-12}$ N. This force can
withstand the attraction only if the charge is $Q\le 10\,\mu$Coul.
So the total energy of the capacitor is of the order of 1 J, which
is considerably smaller of the total energy. But this is true for
a single electron. For a negative ion the potential (\ref{b91})
can be two order of magnitude higher, and it increases $Q$ and its
electrostatic energy.

To create the ball lightning it is necessary to make a point explosion inside
a medium, where the shock wave makes excitation of atoms. Also
it is possible to use an external pumping.
The question is whether the laser discharge will have enough time to be
developed.

To answer this question we compare the time of the light
passage around the ball with that of the shock wave passage over the
distance $d$. The first time is equal to
$T_l=2\pi r_0/c\approx 10^{-9}$ s.
The second one, $T_s$, is defined by an automodel solution~\cite{kor}:
\begin{equation}\label{m5}
r=(t^2W/\rho)^{1/5},
\end{equation}
where $\rho$ is the density of the atmosphere and $W$ is the
energy of the explosion. The speed of the front is equal to
\begin{equation}\label{m7}
v=dr/dt=(2/5)r/t=(2/5)(W/\rho)^{1/2}r^{-3/2}.
\end{equation}
At $W=10^4$ J, $\rho=1$ kg/m$^3$, $r=0,1$ m the speed is $v\approx
10^3$ m/s. So the $T_s\approx 10^{-6}$ s. It shows that $T_l\ll
T_s$, therefore the laser discharge has enough time to be
developed.

It is not necessary that each point explosion will lead to a ball
lightning. Probability of the ball creation is proportional to the
probability of emitting a photon in the WGM mode.

It is interesting that external excited atoms incident on the ball may
be reflected or deexcited. The last channel is the most probable. After
deexcitation the photon layer pulls the atom inside it, so the ball moves in
the direction of the positive gradient of the density of excited atoms and
eats them up.

In fact, to have a stable ball it is not necessary that all
photons inside the ball skin must be coherent. The coherence is
necessary for fast process. If time scale is large enough we can
obtain a similar object with incoherent radiation. It is possible
that the origin of hurricanes and tornado can be related to
similar processes (see also~\cite{ala}).

Till now we considered the scalar case.
A spherical solution for vector electromagnetic field~\cite{akh} in a
reference frame moving with a small velocity $k$ can be represented, for
instance,  in the form
\begin{equation}\label{m8}
{\bf E}=C\exp(i{\bf kr}-i\omega t)\times$$
$$\times\left[\sqrt{\frac{l}{2l+1}}
{\bf Y}_{l,l+1,M}({\bf r}')j_{l+1}(s|{\bf r-k}t|)+
\sqrt{\frac{l+1}{2l+1}} {\bf Y}_{l,l-1,M}({\bf r}')j_{l-1}(s|{\bf
r-k}t|)\right],
\end{equation}
similarly to nonspreading wave packet in quantum
mechanics~\cite{deb}, where ${\bf Y}_{j,l,M}({\bf r}')$ is a
vector spherical harmonics, and ${\bf r}'=({\bf r-k}t)/|{\bf
r-k}t|$.

I am very grateful to E.Shabalin  and V.Yukalov for their moral
support of the presented here idea, and to F.V.Ignatovich, who
read this paper through and found some errors, so I had an
opportunity to correct them in this version.

\end{document}